\documentclass[aps,prd,twocolumn,superscriptaddress,nofootinbib,showpacs,preprintnumbers]{revtex4}
\usepackage[english]{babel}
\begin{document}
\preprint{IMAFF-RCA-07-09}

\title{On the formalism of dark energy accretion onto black- and worm-holes.}

\author{Prado Mart\'{\i}n-Moruno}
\email{pra@imaff.cfmac.csic.es}

\affiliation{Colina de los Chopos, Instituto de Matem\'aticas y
F\'\i sica Fundamental, \\ Consejo Superior de Investigaciones
Cient\'\i ficas, Serrano 121, 28006 Madrid, Spain}

\date{\today}

\begin{abstract}
In this work a general formalism for the accretion of dark energy
onto astronomical objects, black holes and wormholes, is
considered. It is shown that in models with four dimensions or
more, any singularity with a divergence in the Hubble parameter
may be avoided by a big trip, if it is assumed that there is no
coupling between the bulk and this accreting object. If this is
not the case in more than four dimensions, the evolution of the
cosmological object depends on the particular model.
\end{abstract}

 \maketitle

In recent years several cosmological models have been taken into
account in order to explain the current accelerated expansion of
our Universe \cite{Copeland:2006wr}. Some of these models entail
the consideration of new phenomena which may decisively influence
the evolution of the universe. The most popular of them is the
so-called big rip singularity \cite{Caldwell:2003vq}, which is a
possible doomsday for the Universe where both its size and its
energy density become infinitely large. This big rip seemed to be
related to the assumption that the unknown energy density which
carries the accelerated expansion is phantom energy
\cite{Caldwell:1999ew}, i. e., it appeared in models filled with a
fluid with equation of state $p=w\rho$ and $w<-1$. Fortunately, an
escaping route from this doomsday was found, because in these
models there are wormholes which reach an infinite size before the
singularity, \cite{GonzalezDiaz:2004vv}.

The consideration of a particular phantom generalised Chaplygin
gas model (PGCG) in order to describe the accelerating expansion
of the Universe showed that a fluid with phantom nature does not
necessarily imply a big rip singularity
\cite{BouhmadiLopez:2004me} or a big trip \cite{Jimenez
Madrid:2005gd}. However it has been argued that the Universe is
not totally save from a catastrophic doomsday. In fact, another
type of future singularity could take place. It was dubbed the big
freeze singularity and entails that the energy density of the
universe blows up at a finite scale factor. This singularity is
present in some PGCG and some dual phantom generalised Chaplygin
gas (DPGCG) models (the latters can be defined in a
Randall-Sundrum type 1 scenario)
\cite{BouhmadiLopez:2006fu,BouhmadiLopez:2007qb}. Since in the
models which show a big rip singularity, this doomsday can also be
avoided by a big trip, the question is whether this new
singularity, too, associates an extreme growth of wormholes. It is
the aim of the present report to develop a generalized formalism
to describe the dark energy accretion process for black holes and
wormholes, independent of the used cosmic model and of
dimensionality of the espacetime.

\section{Accretion process in four dimensional models.}
It is well known that black holes accrete dark energy in a
different way as ordinary matter does. To study this process, one
must consider that the dark energy density covers the whole space
in a homogeneous and isotropic form. The black hole mass growth
rate for an asymptotic observer can be expressed as
\cite{Babichev:2004yx,MartinMoruno:2006mi}
\begin{equation}\label{bh}
\dot{M}=4\pi D M^2(p+\rho),
\end{equation}
where $D$ is a constant of order unity, the overhead dot denotes
the derivative with respect the cosmic time and $G=c=1$.

On the other hand, it has been shown that phantom models, which
could also describe the current accelerating expansions of the
universe, contain wormholes \cite{Sushkov:2005kj}. Phantom models
are characterized by an equation of state parameter $w<-1$.
Therefore, the accretion process onto wormholes has been also
studied and a similar expression for its rate of variation of
mass, has been obtained \cite{GonzalezDiaz:2004vv},
\begin{equation}\label{wh}
\dot{m}=-4\pi Qm^2(p+\rho),
\end{equation}
with $Q$ a positive constant.

The expansion of a four dimensional homogeneous and isotropic
universe can be described by the Friedmann equations
\begin{equation}\label{F4D}
H^2=\frac{8\pi}{3}\rho
\end{equation}
and
\begin{equation}\label{F4D2}
\frac{\ddot{a}}{a}=-\frac{4\pi}{3}(\rho+3p).
\end{equation}
On the other hand, the conservation law for a perfect fluid is
\begin{equation}\label{conserv}
\dot{\rho}+3H(p+\rho)=0,
\end{equation}
which, taking into account Eq.(\ref{F4D}), yields
\begin{equation}\label{conserv4D}
p+\rho=-\frac{1}{\sqrt{24\pi}}\rho^{-1/2}\dot{\rho}.
\end{equation}
Then, for black holes, Eq.(\ref{bh}) can be expressed as
\begin{equation}
\dot{M}=-\sqrt{\frac{2\pi}{3}}DM^2\rho^{-1/2}\dot{\rho}.
\end{equation}
One can easily integrate this expression and obtain
\begin{equation}\label{M4D}
M=\frac{M_0}{1+\sqrt{\frac{8\pi}{3}}DM_0\left(\rho^{1/2}-\rho_0^{1/2}\right)},
\end{equation}
where the subscript ``0'' denotes current value. Following a
similar pattern for wormholes, one gets
\begin{equation}\label{m4D}
m=\frac{m_0}{1-\sqrt{\frac{8\pi}{3}}Qm_0\left(\rho^{1/2}-\rho_0^{1/2}\right)}.
\end{equation}
This expression was already obtaining for the case in which the
perfect fluid is a particular phantom generalised Chaplygin gas
(PGCG) \cite{Jimenez Madrid:2005gd}, though in the present case
Eq.(\ref{m4D}), Eq.(\ref{M4D}) also does, is valid for any
homogeneous and isotropic model in four dimensions. Taking into
account Eq.(\ref{F4D}), both mass functions can be expressed in
terms of the Hubble parameter \cite{Yurov:2006we}
\begin{equation}\label{M4DH}
M(t)=\frac{M_0}{1+DM_0\left[H(t)-H_0\right]}
\end{equation}
and
\begin{equation}\label{m4DH}
m(t)=\frac{m_0}{1-Qm_0\left[H(t)-H_0\right]}.
\end{equation}
Thus, the black hole mass will increase if and only if the Hubble
parameter decreases. Then, since
$\dot{H}=\ddot{a}/a-H^2=-4\pi(p+\rho)$, $p+\rho>0$, a conclusion
which can also be directly extracted from Eq.(\ref{bh}). It
follows that in order to infinitely increase the mass in a finite
time, the black hole must have a current mass $M_0>1/(DH_0)\sim
10^{23}M_\odot$, and that is not possible
\cite{MartinMoruno:2006mi}. This mass will decrease in phantom
models, $p+\rho<0$, vanishing where $H(t)$ diverges, in case that
this eventually happens. On the other hand, a similar argument
shows that the wormhole mass will increases only in phantom
models, where the wormhole throat became infinitely large at
$t_*$, with $H(t_*)=H_0+1/(Qm_0)<\infty$, if the universe reaches
that time. But, if there is a singularity in the model in which
the Hubble parameter is infinity, since the Hubble parameter is an
ever increasing function, then $t_0<t_*<t_{{\rm sing}}$. Therefore
the wormhole would have an infinitely large size before the
universe reaches the singularity, that singularity occurring at a
finite or infinite scale factor, i. e., a big trip could avoid all
the singularities with an infinite Hubble parameter that could
appear in any homogeneous and isotropic four dimensional model.

\vspace{12pt}

Now we consider the particular case of phantom generalised
Chaplygin gas models. As it is well known, generalised Chaplygin
gas (GCG) satisfies the equation of state \cite{Kamenshchik}
\begin{equation}\label{uno}
p=-\frac{A}{\rho^{\alpha}},
\end{equation}
where $A$ is a constant and $\alpha$ is a parameter (corresponding
to the case $\alpha=1$ in the original Chaplygin gas). The
conservation law for the energy momentum tensor for this fluid can
be expressed as
\begin{equation}\label{dos}
\rho=\left(A+\frac{B}{a^{3(1+\alpha)}}\right)^{\frac{1}{1+\alpha}},
\end{equation}
with $B$ a constant (which can be expressed in terms of $A$,
$\rho_0$ and $a_0$). Such a fluid must fulfill $\rho>0$ and
$p+\rho<0$ in order to have a phantom nature,
\cite{BouhmadiLopez:2004me,Khalatnikov:2003im}. It can be shown
\cite{BouhmadiLopez:2007qb}, that there are four different kinds
of phantom generalised Chaplygin gas (PGCG) models, corresponding
to four different universes. The cosmological scenarios fulfilling
$1+\alpha>0$ are asymptotically de Sitter in the future. The two
models with $1+\alpha<0$, show a big freeze singularity at a
finite time in the future, i. e., the scale factor grows to a
maximum finite value where the energy density diverges. These
singular models correspond to $A>0$, $B<0$ and $1+\alpha<0$, and
$A<0$, $B>0$ and $(1+\alpha)^{-1}=2n<0$ (with $n$ some negative
integer number), respectively, and the maximum scale factor in
both cases is
\begin{equation}
a_{{\rm max}}=\left|\frac{B}{A}\right|^{\frac{1}{3(1+\alpha)}}.
\end{equation}

As this singularity implies a divergence on the Hubble parameter a
big trip phenomenon should take place, as we have argued before.
In fact, if one defines $x=a/a_{{\rm max}}$ ($0\leq x\leq1$),
Eq.(\ref{m4DH}) for the first mentioned case can be re-expressed
as
\begin{equation}\label{masachap}
m(x)=\frac{m_0}{1+\sqrt{\frac{8\pi}{3}}\frac{Qm_0}{|A|^{n}}\left[\frac{1}{(1-x_0^{3p})^{n}}-\frac{1}{(1-x^{3p})^{n}}\right]}.
\end{equation}
It is the same for the second considered case replacing $|A|$ for
$A$ and $n$ for $-1/[2(1+\alpha)]$. Since $m(0)<m_0$ and
$m_0>m(1)$, either the function $m(x)$ changes from an increasing
function to a decreasing function in the x-interval $(0,1)$, or it
diverges at least once on this interval. If one defines
$F(x)=m_0/m(x)$, it is easy to see that $F(x)$ is a continuous and
increasing function in the whole interior of this interval.
Therefore $m(x)$ is an increasing function with $m_0>m(1)$, so it
must diverge in $(x_0,1)$. Then the wormhole mass and throat size
will be infinitely large before the universe reaches its doomsday
(at $x=1$), i.e., these universes present a big trip which could
avoid the big freeze singularity as it was expected.

\section{Accretion in higher dimensional models.}
If we consider cosmological models constructed in a space with
more than four dimensions, the Friedmann equation would read
\begin{equation}\label{F5D}
H^2=\frac{8\pi}{3}F(\rho),
\end{equation}
where $F(\rho)$ is a generic function of $\rho$, $\rho$ being the
dark energy density which is defined on the bulk space. It is well
known that the correction introduced in Eq.(\ref{F4D}) which leads
to Eq.(\ref{F5D}) must be relevant only at hight energies
\cite{Cline:1999ts}. Since in models with $w+1<0$ the energy
density increases with time, this correction could even be
dominant at the late time acceleration period. Moreover, in this
work we consider dual models which imply an absolute value of the
energy density sufficiently large to do this correction
indispensable.

As we only see four dimensions, we could treat that as if the
expansion of our universe was originated by a effective
$\rho_{eff}=F(\rho)$, which is the projection of $\rho$ onto four
dimensions. Therefore, in order to consider the accretion process
there are two possibilities. Firstly, when the astronomical objet
can accrete the whole fluid on the bulk or, secondly, when the
object placed in the four dimensional patch only accretes the
effective fluid which can be defined on these dimensions.

\subsection{Accretion of the fluid on the bulk.}
Here we will assume that there is some coupling between the bulk
and the brane, allowing the astronomical object to accrete the
fluid which drives the physics of the universe, as it is done in
\cite{Yurov:2006we} for a particular model. Considering
Eq.(\ref{F5D}) and the conservation law Eq.(\ref{conserv}), we can
obtain
\begin{equation}
p+\rho=-\frac{1}{\sqrt{24\pi}}\rho_{eff}^{-1/2}\dot{\rho},
\end{equation}
and inserting this expression in Eq. (\ref{bh})
\begin{equation}
\frac{{\rm d}M}{M^2}=-\sqrt{\frac{2\pi}{3}}D\rho_{eff}^{-1/2}{\rm
d}\rho.
\end{equation}
In order to integrate this expression one must know the specific
functional form of $F(\rho)$. Therefore the black hole mass
function will depend on the considered cosmological model and the
same conclusion can be extracted in the wormhole case.

A Randall-Sundrum type 1 scenario, \cite{RS}, is a five
dimensional model with a four dimensional brane which could
represent our Universe. The scale factor in the four dimensional
universe can be defined with the modified from Friedmann equation
on the brane
\begin{equation}\label{MFE}
H^2=\frac{8\pi}{3}\rho\left(1+\frac{\rho}{2\lambda}\right),
\end{equation}
where $\lambda$ is the positive brane tension and $\rho$ the
energy density on the bulk. In this scenario, one can consider a
phantom fluid, $w<-1$, which fulfill the null energy condition,
$p+\rho>0$, and has a well defined Hubble parameter; this fluid is
known as dual phantom fluid \cite{Yurov:2006we}. That would
implied an energy density negative definite, with
$\rho<-2\lambda$, but it must be also noticed that the effective
energy density on the brane keeps always being positive.

A dual phantom generalised Chaplygin gas (DPGCG) is defined as a
perfect fluid satisfying Eq.(\ref{dos}), with the above-mentioned
characteristics \cite{BouhmadiLopez:2006fu,BouhmadiLopez:2007qb}.
These characteristics are satisfied provided that $A<0$ and $B>0$
in Eq.(\ref{dos}). If one also considers $(1+\alpha)^{-1}=2n+1$
with $n$ a negative integer, two models are obtained which show a
big freeze singularity in the future. In these models
Eqs.(\ref{uno}), (\ref{dos}) and (\ref{MFE}) imply
\begin{equation}\label{pmasrho}
p+\rho=\frac{x^{\frac{3}{1+2p}}}{|A|^{1+2p}\left(1-x^{\frac{3}{1+2p}}\right)^{2(1+p)}},
\end{equation}
where $1+2p=-(1+2n)$ with $x<1$ and no-negative ($x=1$
corresponding to the big freeze). Replacing Eq.(\ref{dos}) in Eq.
(\ref{MFE}), one obtains
\begin{equation}\label{derivadax}
\dot{x}=\sqrt{\frac{8\pi}{3}}x\frac{\left[1-2\lambda
|A|^{1+2p}\left(1-x^{\frac{3}{1+2p}}\right)^{1+2p}\right]^{1/2}}{\sqrt{2\lambda}|A|^{1+2p}\left(1-x^{\frac{3}{1+2p}}\right)^{1+2p}}.
\end{equation}

Taking into account Eqs.(\ref{pmasrho}) and (\ref{derivadax}), one
can integrate Eq.(\ref{wh}) to yield
\begin{equation}
m(k)=\frac{m_0}{1+\sqrt{\frac{4\pi\lambda}{3}}Q m_0\left[{\rm
Ln}\left(\frac{1+k}{1-k}\right)-{\rm
Ln}\left(\frac{1+k_0}{1-k_0}\right)\right]},
\end{equation}
with $k=\left[1-2\lambda
|A|^{1+2p}\left(1-x^{\frac{3}{1+2p}}\right)^{1+2p}\right]^{1/2}$.

Therefore, when $a$ increases, $k$ increases and viceversa. The
${\rm Ln}\left[(1+k)/(1-k)\right]$ is a positive increasing
function in the whole interval, because $(1+k)/(1-k)>1$ for these
values. This implies that the wormhole mass will decrease when the
scale factor increases; in fact, replacing the value $x=1$ in the
above expression, one can see that the wormhole tends to disappear
at the singularity.

On the other hand, it is shown that in some dual phantom models
\cite{Yurov:2006we} a big hole phenomenon could take place, i. e.,
a black hole could grow up so rapidly that its size could become
infinitely large before the singularity. To see if that is the
case we must integrate the black hole mass rate Eq.(\ref{bh}) for
an asymptotic observer. In these models this integration produces
\begin{equation}\label{Mbulk}
M=\frac{M_0}{1+\sqrt{\frac{4\pi\lambda}{3}}DM_0\left[{\rm
Ln}\left(\frac{1+k_0}{1-k_0}\right)-{\rm
Ln}\left(\frac{1+k}{1-k}\right)\right]}.
\end{equation}
The Ln takes values $0\leq {\rm Ln}\left[(1+k)/(1-k)\right]\leq
\infty$ for $k$ in $[0,1]$, so there must be some $k_0<k_*<1$ for
which the denominator of Eq.(\ref{Mbulk}) vanishes; so this
expression diverges before the universe reaches its maximum size.
Therefore a black hole would increase its size so rapidly that
would eventually engulf the universe itself, which always has a
finite size, before it reaches the singularity.

\subsection{Accretion of the fluid on the brane.}

If the considered object accretes the fluid that is defined in the
same dimensions as it, as studied in \cite{Zhang:2007yu} for a
particular case, then in Eq.(\ref{bh}) and (\ref{wh}) we must have
$\rho_{eff}+p_{eff}$. Differentiating Eq.(\ref{F5D}) one obtains
\begin{equation}\label{dH}
2H\dot{H}=\frac{8\pi}{3}\frac{{\rm d}\rho_{eff}}{{\rm
d}\rho}\dot{\rho},
\end{equation}
then, using the conservation law given by Eq.(\ref{conserv}),
\begin{equation}
\dot{H}=-4\pi\frac{{\rm d}\rho_{eff}}{{\rm d}\rho}(p+\rho).
\end{equation}
In order to define an effective pressure, we take
\begin{equation}
\frac{\ddot{a}}{a}=\dot{H}+H^2=-\frac{4\pi}{3}(\rho_{eff}+3p_{eff}),
\end{equation}
which implies
\begin{equation}
p_{eff}+\rho_{eff}=(p+\rho)\frac{{\rm d}\rho_{eff}}{{\rm d}\rho}.
\end{equation}
Now Eq. (\ref{bh}) can be expressed as
\begin{equation}
\frac{{\rm d}M}{M^2}=4\pi D(p+\rho)\frac{{\rm d}\rho_{eff}}{{\rm
d}\rho}{\rm d}t,
\end{equation}
which, taking into account Eq.(\ref{conserv}) and (\ref{F5D}),
yields
\begin{equation}
\frac{{\rm d}M}{M^2}=-\sqrt{\frac{2\pi}{3}}D\rho_{eff}^{-1/2}{\rm
d}\rho_{eff}.
\end{equation}
This expression can be easily integrated to obtain the black hole
mass
\begin{equation}
M=\frac{M_0}{1+\sqrt{\frac{8\pi}{3}}DM_0\left(\rho_{eff}^{1/2}-\rho_{eff0}^{1/2}\right)},
\end{equation}
and following a similar way
\begin{equation}\label{m5D}
m=\frac{m_0}{1-\sqrt{\frac{8\pi
}{3}}Qm_0\left(\rho_{eff}^{1/2}-\rho_{eff0}^{1/2}\right)},
\end{equation}
for the wormhole mass. We must also point out that these
expressions are the same as Eq.(\ref{M4D}), replacing $\rho$ by
$\rho_{eff}$, the energy density defined in our universe. In fact,
one can easily see that in terms of the Hubble parameter one
obtains the same expressions as in the four dimensional case
\begin{equation}
M=\frac{M_0}{1+DM_0\left[H(t)-H_0\right]}
\end{equation}
and
\begin{equation}\label{m5DH}
m=\frac{m_0}{1-Qm_0\left[H(t)-H_0\right]}.
\end{equation}
As in section I, the black hole mass is an increasing function if
the Hubble parameter is a decreasing function, but now that occurs
if $p+\rho>0$ and ${\rm d}\rho_{eff}/{\rm d}\rho>0$ or if
$p+\rho<0$ and ${\rm d}\rho_{eff}/{\rm d}\rho<0$, both cases
corresponding to an effective dark energy fluid with $w_{eff}>-1$.
A big hole phenomenon is not possible because, as it was mentioned
before, there can not be black holes with the necessary initial
mass in our Universe. On the other hand, wormholes will increase
if the effective fluid is a phantom fluid, which occurs if
$p+\rho>0$ and ${\rm d}\rho_{eff}/{\rm d}\rho<0$ or if $p+\rho<0$
and ${\rm d}\rho_{eff}/{\rm d}\rho>0$. In both cases the fluid
will be phantom or not on the bulk depending on the specific
scenario. If in the model there is a singularity where the Hubble
parameter tends to infinity, the black holes would disappear at
this singularity. And because $H_0$ is finite and the Hubble
parameter is a continuous function, there must necessarily be a
time $t_0<t_*<t_{sing}$ defined for $H(t_*)=H_0+1/(Qm_0)$, where
the wormhole will become infinitely large; therefore the universe
does not reach the singularity because a big trip phenomenon would
take place before.

In a Randall-Sundrum type I scenario, the effective energy density
and pressure on the brane can be defined
\begin{equation}
\rho_{eff}=\rho\left(1+\frac{\rho}{2\lambda}\right)
\end{equation}
and
\begin{equation}
p_{eff}=p\left(1+\frac{\rho}{\lambda}\right)+\frac{\rho^2}{2\lambda},
\end{equation}
so that $w_{eff}=p_{eff}/\rho_{eff}$, and one has
\begin{equation}\label{weff}
1+w_{eff}=(1+w)\frac{1+\rho/\lambda}{1+\rho/(2\lambda)}.
\end{equation}
Therefore, for this scenario with an arbitrary homogeneus and
isotropic fluid on the bulk (not necessarily a generalised
Chaplygin gas), the sign of $1+w_{eff}$ is the same as the sign of
$1+w$. Then in the DPGCG models described in the previous
subsection, we have an effective phantom fluid in the brane with,
obviously, $\rho_{eff}>0$. Moreover, as the fraction in
Eq.(\ref{weff}) is always positive and bigger than 2, $1+w_{eff}$
is more negative than $1+w$, i. e., it is ''more phantom''.

In the models mentioned above, Eq.(\ref{m5DH}) yields
\begin{equation}
m(k)=\frac{m_0}{1+\sqrt{\frac{16\pi\lambda}{3}}Qm_0\left(\frac{k_0}{1-k_0^2}-\frac{k}{1-k^2}\right)},
\end{equation}
where $k$ is the same as in the previous subsection. One can
define $f(k)=k/(1-k^2)$. As $f(k_0)<f(k)<\infty$ for $x_0<x<1$,
there is a $k_*<\infty$ (which implies $x_*<1$) with
$f(k_*)=f(k_0)+1$, where the wormhole mass diverges. Therefore, a
big trip would occur as one would expect under general grounds.

\section{Conclusions.}

In this work we have seen that in four dimensional, homogeneous
and isotropic models, black holes grow up if the Hubble parameter
is a decreasing function, which implies $w>-1$; otherwise, they
would shrink down to zero in a finite time in the presence of a
singularity with divergent $H(t)$, or asymptotally if there is no
such a future singularity. It must be also pointed out that in
dark GCG (DGCG) models with $0<\alpha<1$, generalised Chaplygin
dark stars should exist that would increase their size with time
\cite{Bertolami:2005pz}. Since this is a dark model, $w>-1$, black
holes should also grow, as it could be expected from the behaviour
of the dark stars. On the other hand, wormholes would shrink if
$H(t)$ decreases and increase its size if $H(t)$ is an increasing
function. In fact, in models which show a future singularity with
divergence in the Hubble parameter (which in these models means a
divergence in the energy density), this singularity would be
avoided by a big trip, because there is a time $t_*$ before the
singularity where the wormhole would be infinitely large. In
particular, we have shown that this big trip occurs in PGCG models
also with a big freeze.

The case of models with more than four dimensions is a little more
complicated, because it depends on whether the object could
accrete the whole fluid on the bulk or accretes only the effective
fluid which can be defined on the brane where the wormhole is. In
the first case, we have shown that the mass function will depend
on the model considered. In the DPGCG case, wormholes decrease
their size when the scale factor increases, so a big trip
phenomenon is not possible. But, in this case, black holes grow up
and can engulf the universe before the singularity. In the second
case, we consider that the object is able to accrete only the
effective fluid, so that the conclusions are the same as in
four-dimensional models, i. e., any singularity implying a
divergence of the Hubble parameter that could take place at a
finite time in the future is avoided by a big trip. In DPGCG
models the results obtained following the above two ways are the
opposite, since though in the first case the universe avoids a
doomsday, it may find an even more dramatic end. In the second
case the universe embarks itself in a trip with unknown
destination. Besides physical considerations, simplicity and
beauty dictate that, for cosmological models with a modified
Friedmann equation, the accretion process is most appropriately
dealt with by the considering accretion of the effective fluid
with four-dimensional energy density. If we take into account
recent work by Frampton \cite{Frampton:2007cr} who argued that
effective energy density and pressure are not proper physical
quantities, it appears that one ought to choose the former of the
above two interpretations.

\section*{Acknowledgements}
The author thanks Pedro F.~Gonz\'alez D\'{\i}az for useful
discussions and encouragement and Marco Tripodi for valuable
technical help and creating a wonderful working atmosphere. This
work was supported by MEC under Research Project No.FIS2005-01181.
The author gratefully acknowledges the financial support provided
by the I3P framework of CSIC and the European Social Fund.

\end{document}